\documentclass[pdflatex,sn-mathphys-num]{sn-jnl}

\usepackage[T1]{fontenc}
\usepackage[utf8]{inputenc}
\usepackage{lineno,hyperref}
\modulolinenumbers[20]
\usepackage{float}
\usepackage{graphicx, subfigure}
\usepackage{color}
\usepackage{tcolorbox}
\usepackage{booktabs}
\usepackage{longtable}
\usepackage{amssymb}
\usepackage{amsmath}
\usepackage{dsfont}
\usepackage{bbm}
\usepackage{lipsum}
\usepackage{enumitem}
\usepackage{multirow}
\usepackage{listings}
\usepackage[table]{xcolor}

\usepackage{listings}
\usepackage{xcolor}

\usepackage{wasysym}
\usepackage{algorithm}
\usepackage{algpseudocode}

\lstdefinelanguage{nasm}{
  morekeywords={mov,xor,add,sub,inc,dec,cmp,jmp,jl,jle,jg,jge,je,jne,jz,jnz,
                push,pop,call,ret,nop,lea,test,and,or,shl,shr,imul,idiv,
                short,near,far,byte,word,dword,qword,ptr,ds,cs,ss,es,fs,gs,
                section,global,extern,db,dw,dd,dq},
  morekeywords=[2]{eax,ebx,ecx,edx,esi,edi,ebp,esp,
                   rax,rbx,rcx,rdx,rsi,rdi,rbp,rsp,
                   ax,bx,cx,dx,al,bl,cl,dl,ah,bh,ch,dh},
  sensitive=false,
  morecomment=[l]{;},
  morestring=[b]",
  morestring=[b]',
  alsodigit={0x}
}

\lstdefinestyle{nasmstyle}{
  language=nasm,
  basicstyle=\ttfamily\scriptsize,
  keywordstyle=\color{blue!70!black}\bfseries,
  keywordstyle=[2]\color{purple!80!black},
  commentstyle=\color{gray}\itshape,
  stringstyle=\color{orange!80!black},
  numberstyle=\tiny\color{gray},
  frame=single,
  rulecolor=\color{black!50},
  showstringspaces=false,
  breaklines=true,
  columns=fullflexible,
  keepspaces=true,
  tabsize=2,
  xleftmargin=2pt,
  xrightmargin=2pt
}

\lstdefinestyle{cstyle}{
  language=C,
  basicstyle=\ttfamily\scriptsize,
  keywordstyle=\color{blue!70!black}\bfseries,
  commentstyle=\color{gray}\itshape,
  stringstyle=\color{orange!80!black},
  numberstyle=\tiny\color{gray},
  frame=single,
  rulecolor=\color{black!50},
  showstringspaces=false,
  breaklines=true,
  columns=fullflexible,
  keepspaces=true,
  tabsize=2,
  xleftmargin=2pt,
  xrightmargin=2pt
}

\usepackage{tikz}
\usepackage{adjustbox}
\usepackage{amssymb}
\usetikzlibrary{shapes.geometric, arrows.meta, positioning, fit, backgrounds, shadows, calc, patterns, decorations.pathmorphing}

\definecolor{techblue}{RGB}{230, 240, 250}       
\definecolor{techborder}{RGB}{70, 130, 180}      
\definecolor{processyellow}{RGB}{255, 242, 204}  
\definecolor{securegreen}{RGB}{200, 230, 200}    
\definecolor{alertred}{RGB}{245, 200, 200}       
\definecolor{darkslate}{RGB}{50, 50, 60}         
\definecolor{softgray}{RGB}{245, 245, 245}       

\usepackage[numbers]{natbib}

\graphicspath{{figs/}}
\newcommand{\sysname}{\texttt{MAGMA}~}

\begin{document}

\title[MAGMA]{Quantifiable Uncertainty: A Stochastic Consensus Multi-Agent RAG Framework for Robust Malware Detection}

\author*[1]{\fnm{ElMouatez Billah} \sur{Karbab} }\email{mouatez@karbab.net}
\affil*[1]{\orgname{Joaan Bin Jassim Academy for Defence Studies}, \orgaddress{\city{Al-Khor}, \country{Qatar}}}



\abstract{
While contemporary deep learning malware detectors define a dominant defense paradigm, their sophistication also exposes them to novel structural evasion attacks, a limitation we attribute to their inherent inability to express epistemic uncertainty. To address this challenge, we present \sysname, a Retrieval-Augmented Generation (RAG) framework that decouples malware analysis into semantic code retrieval and probabilistic verification. In contrast to monolithic classifiers, \sysname\ employs a dual-stream embedding scheme over assembly and pseudo-code representations to isolate \textit{Decision-Critical Functions} (DCFs) from the noise of dead code. We further introduce a \textbf{Stochastic Consistency Ensemble}, in which multiple instances of the same reasoning agent independently evaluate the retrieval set under non-deterministic sampling. From this ensemble, we derive two complementary metrics: \textbf{Function Evidence Strength (FES)}, a weighted aggregation of retrieval confidence, and the \textbf{Evidence Conflict Score (ECS)}, defined as the Shannon entropy of the ensemble's predictive distribution. We show that elevated ECS values serve as an effective proxy for structural ambiguity, enabling the system to implement a principled ``reject-option'' policy. Extensive evaluation demonstrates that \sysname\ achieves a 98.4\% detection rate, substantially exceeding existing solutions.
}


\maketitle
\section{Introduction}
Malware defense operates under a fundamental asymmetry. Defenders face a strict \textit{soundness constraint}, requiring them to consistently bound all malicious variants. Evasive threats, conversely, operate under an \textit{existential sufficiency}, needing only a single successful evasion path to bypass detection entirely \cite{Anderson_Ross2001Why}. The proliferation of automated structural mutation tools and metamorphic engines widens this gap. These systems apply diverse techniques to generate a theoretically infinite number of polymorphic variants from a single malicious payload.

Traditional static analysis relies on fixed signature matching (e.g., YARA rules~\cite{Li2023PackGenome}) and struggles to detect these novel permutations. Dynamic analysis, or sandboxing, provides higher fidelity but incurs prohibitive computational costs. It also remains vulnerable to red-pill environmental checks, a scenario where malware detects the virtualized environment and remains dormant \cite{Bulazel_Alexei2017Survey}. The security industry has thus shifted toward Deep Learning (DL) based static analysis frameworks \cite{GibertDaniel2020Rise}. The current generation of DL-based detectors, however, introduces a critical vulnerability known as \textbf{semantic blindness}. Because these models process binary code as purely statistical data without semantic reasoning, they act as stochastic pattern matchers. Subtle structural perturbations can shift the feature space representation across a decision boundary while preserving the underlying malicious functionality \cite{Arp_Daniel2022Dos}.

\subsection{Problem Statement}
\label{sec:intro_limitations}
Deterministic classifiers dominate neural malware detection but introduce structural vulnerabilities by outputting a single point-estimate verdict $P(y|x)$. In security-critical contexts, this design flaw manifests as \textbf{Confident Wrongness}~\cite{Hein_Matthias2019Why}. Standard Softmax outputs measure distance to a decision boundary rather than the epistemic certainty of the model itself~\cite{Gal_Yarin2016Dropout}.

Suppose an evasion engine $\mathcal{E}$ applies a transformation $T(x) \to x'$ such that $x'$ retains the malicious semantics of $x$ while occupying a distinct region in the embedding space. A deterministic model $M$ will often classify $M(x')$ as benign with high confidence ($>99\%$) simply because the transformed input falls into a sparse region of the high-dimensional manifold. State-of-the-art systems~\cite{HeYuewang2024ResNeXt} lack a \textbf{formal rejection mechanism}. They force a binary decision even when the input lies outside the training data distribution. A robust security system must quantify its \textit{epistemic uncertainty}. It must identify its own blind spots and flag anomalous cases for human review rather than committing a catastrophic automated error.

\subsection{Proposed System}
\label{sec:intro_system}
We propose \sysname (Multi-Agent Generative Malware Analysis), a framework that reformulates detection from static pattern-matching into a \textit{probabilistic evidence-gathering process}.

The architecture centers on \textit{Retrieval-Augmented Generation (RAG)}, bridging the gap between the frozen parametric knowledge of Large Language Models (LLMs) and the rapidly evolving threat landscape. The pipeline employs a \textit{Dual-Stream} lifting mechanism that processes two parallel representations: raw assembly instructions for syntactic structure and decompiled pseudo-code for semantic control flow.

Traditional classifiers compress features into a single decision vector. \sysname instead projects these dual representations into a joint vector space and queries a dynamic \textit{Vector Knowledge Base}. This retrieval step isolates the $k$-nearest \textbf{Decision-Critical Functions (DCFs)}. We define DCFs as verified snippets of malicious logic extracted from high-confidence malware samples. Grounding downstream reasoning in retrieved evidence prevents hallucinations. The system detects novel variants by matching semantic similarities to known logic blocks without requiring model retraining.

\sysname evaluates these retrieved contexts using a stochastic multi-agent ensemble. Instead of relying on a single deterministic inference, the framework aggregates reasoning from diverse agents specialized in structural similarity, malicious intent, and attribution. This consensus computes a weighted \textbf{Function Evidence Strength ($FES$)}. The system also calculates an \textbf{Evidence Conflict Score ($ECS$)} to measure inter-agent disagreement. The final output is not a mere classification label but a tuple $\langle \text{Verdict}, FES, ECS \rangle$. This structure allows the system to autonomously reject ambiguous inputs when structural perturbations maximize epistemic uncertainty.

\subsection{Contributions}
\label{sec:intro_contributions}

\begin{enumerate}
    \item \textbf{Dual-Stream RAG Architecture:} A retrieval mechanism maps syntactic assembly embeddings with semantic pseudo-code logic. Grounding inference in $k$-nearest neighbors from a curated Knowledge Base of Decision-Critical Functions (DCFs) mitigates the vulnerabilities of opaque neural networks.

    \item \textbf{Quantifiable Attribution via Evidence Strength:} We introduce the \textit{Function Evidence Strength} ($FES$) metric to evaluate retrieved contexts. By weighting the influence of each DCF, $FES$ quantifies how specific code segments affect the final verdict. This transforms black-box predictions into auditable evidence chains.

    \item \textbf{Stochastic Self-Consistency via Multi-Agent Consensus:} The system models malware detection as a probabilistic consensus game. \sysname triggers a stochastic debate across an ensemble of reasoning agents ($T > 0$) rather than executing a single deterministic pass. These agents independently evaluate retrieved evidence against adversarial criteria to produce a bounded predictive distribution.

    \item \textbf{Formalized Rejection via Entropy:} We define the \textit{Evidence Conflict Score} ($ECS$) as the Shannon entropy ($H(X)$) of the ensemble's verdict distribution. $ECS$ serves as a proxy for sensitivity to adversarial perturbations. An $ECS$ exceeding a defined threshold triggers a \textbf{fail-safe rejection}, stopping the system from rendering classifications when the reasoning engine encounters severe adversarial noise.

    \item \textbf{Extensive Evaluation:} We test the system against highly polymorphic malware variants. \sysname outperforms established static analysis baselines (MalConv~\cite{RaffEdward2018Malware} and ResNeXt+~\cite{HeYuewang2024ResNeXt}). Enforcing the entropy-guided rejection policy yields a 4.2\% absolute improvement in detection accuracy against evasion attempts alongside a 4.0\% reduction in the False Positive Rate (FPR).
\end{enumerate}


\section{Background \& Threat Model}
\label{sec:threat_model}

\subsection{Malware Representation \& Lifting}
\label{sec:bg_representation}
In the context of malicious code identification, we formalize the malware detection task as a binary classification problem over the set of executable binaries $\mathcal{X} = \{0, 1\}^*$. A classifier acts as a continuous function $f: \mathcal{X} \rightarrow [0, 1]$, representing the calculated probability that a given binary $x \in \mathcal{X}$ executes malicious logic.

Analyzing raw executable files has historically presented distinct structural challenges. Baseline deep learning architectures (e.g., MalConv) typically represent the input $x$ strictly as a flat sequence of raw bytes, $x = \{b_1, b_2, \dots, b_L\}$. While this computational efficiency enables a highly scalable detection mechanism, it fundamentally introduces a vulnerability to \textit{semantic blindness}. By treating the binary as an unstructured sequence, the neural network learns to leverage statistical correlations in byte frequencies rather than reasoning about the actual execution logic. This reliance produces a decision boundary that remains highly brittle to problem-space shifts and structural mutation.

To mitigate this semantic opacity, we define a \textit{Dual-Stream Lifting Function} $L: \mathcal{X} \rightarrow \mathcal{I}_{ASM} \times \mathcal{I}_{Code}$. This framework transforms the opaque binary payload into two distinct, human-readable intermediate representations prior to executing the feature vectorization process:
\begin{itemize}
    \item \textbf{Syntactic Stream ($\mathcal{I}_{ASM}$):} This stream applies a linear disassembly mapping to the binary $x$, producing normalized assembly instructions. This perspective explicitly captures the granular architectural dependencies, precise register manipulations, and cryptographic constants that typically vanish during high-level decompilation.
    \item \textbf{Semantic Stream ($\mathcal{I}_{Code}$):} This representation decompiles the binary $x$ into high-level pseudo-C code. The transformation abstracts away compiler-specific register allocations and stack configurations, isolating the underlying Control Flow Graph (CFG), explicit data dependencies, and core API call sequences.
\end{itemize}

\begin{figure}[h]
\centering

\begin{lstlisting}[style=cstyle]
// *** Semantic Stream (Pseudo-Code) ***
// Clear intent: Opens a network connection
void call_c2() {
    SOCKET s = socket(AF_INET, SOCK_STREAM, 0);
    connect(s, &target_addr, sizeof(target_addr));
}
\end{lstlisting}

\begin{lstlisting}[style=nasmstyle]
; *** Syntactic Stream (Assembly) ***
; Obfuscated structure, but exact register tracking
push ebp
mov  ebp, esp
sub  esp, 0x10
push 0         ; protocol
push 1         ; SOCK_STREAM
push 2         ; AF_INET
call ds:socket ; Indirect API call
\end{lstlisting}

\caption{\textbf{The Semantic Gap.} This paradigm illustrates the distinct dual representations extracted by $inline$L(x)$inline$. The semantic stream distinguishes high-level API intent, while the syntactic stream leverages exact register operations often targeted by evasion engines.}
\label{lst:lifting_example}
\end{figure}

This dual representation ensures that the semantic stream effectively absorbs purely syntactic variations (e.g., instruction substitution or register renaming), while the assembly stream anchors the analysis against high-level logic obfuscations to maintain a resilient framework. 

\begin{figure}[t]
\centering
\includegraphics[width=\textwidth]{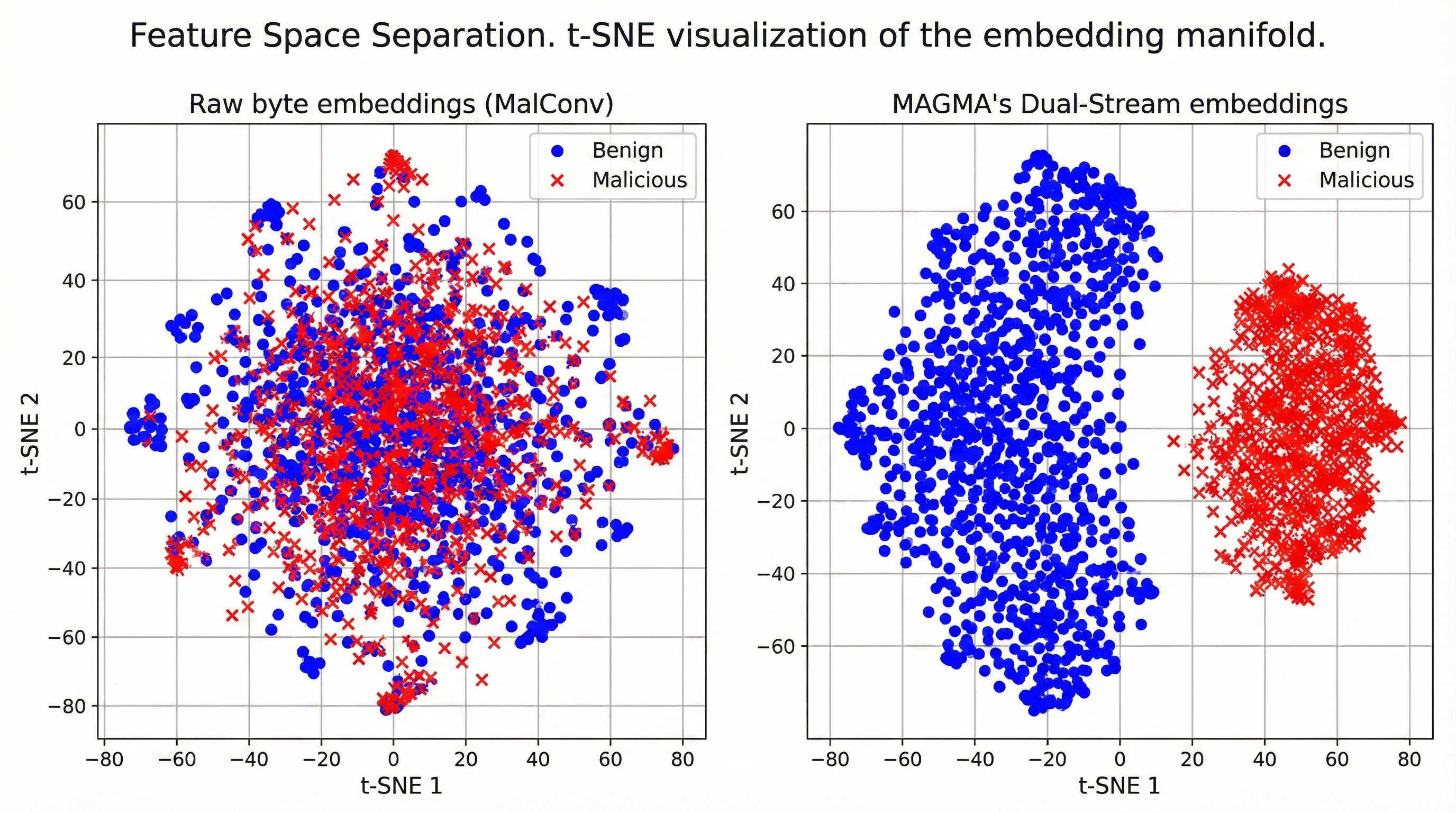}
\caption{\textbf{Feature Space Separation.} t-SNE visualization of the embedding manifold. \textbf{(Left)} Raw byte embeddings (e.g., MalConv) exhibit significant class overlap, attributing their vulnerability to boundary shifting. \textbf{(Right)} The Dual-Stream embeddings of \sysname orchestrate distinct semantic clusters, forcing the evasion engine to leverage larger, resilient problem-space perturbations to cross the boundary.}
\label{fig:tsne}
\end{figure}

Figure \ref{fig:tsne} provides direct empirical evidence for the necessity of this lifting phase. As illustrated in the left panel of Figure \ref{fig:tsne}, raw-byte embeddings produce significant class overlap. Malicious and benign samples occupy the same latent space because they share standard compiler artifacts and library headers. The right panel demonstrates that the Dual-Stream embeddings of \sysname successfully project the binaries into distinct, well-separated semantic clusters. Restructuring the feature space in this manner forces evasion engines to apply drastically larger, functionality-breaking perturbations just to cross the decision boundary.

\subsection{Threat Model and Scope}
\label{sec:threat_model_scope}

To formalize the threat landscape, we model a strategic, Probabilistic Polynomial-Time (PPT) adversary $\mathcal{A}$ operating an evasion engine $\mathcal{E}$ strictly within the \textit{Problem-Space Domain}~\cite{PierazziFabio2020Intriguing}. Our evaluation focuses on realistic scenarios where $\mathcal{A}$ manipulates functional binary structures to maximize evasion probability without altering the underlying malicious semantics. Table~\ref{tab:threat_model} details our formal assumptions regarding trust boundaries, adversarial capabilities, and system scope.

\begin{table}[h]
\centering
\caption{\textbf{Taxonomy of the Threat Model and System Scope.}}
\label{tab:threat_model}
\begin{tabular}{l|p{0.75\columnwidth}}
\toprule
\textbf{Parameter} & \textbf{Formal Definition \& Constraints} \\
\midrule
\textbf{Adversary Profile} & Problem-Space, Probabilistic Polynomial-Time (PPT) attacker. \\
\midrule
\textbf{Knowledge Level} & \textbf{Gray-Box:} Understands architecture. No access to $\mathcal{KB}$, weights ($\theta$), or prompts. \\
\midrule
\textbf{Oracle Access} & Hard-label and confidence score observation via iterative probing $\mathcal{O}(x)$. \\
\midrule
\textbf{Perturbation Space} & Instruction substitution, dead code insertion, and block transposition ($\Delta \leq \epsilon$). \\
\midrule
\textbf{Out-of-Scope} & 
\vspace{-2mm}
\begin{itemize}[leftmargin=*, nosep]
    \item \textit{Dynamic Evasion:} Sandbox/Hypervisor detection (``Red-Pills'').
    \item \textit{Heavy Obfuscation:} Complete payload virtualization (e.g., VMProtect).
    \item \textit{TCB Compromise:} Poisoning of the Vector Knowledge Base ($\mathcal{KB}$).
\end{itemize} \\
\bottomrule
\end{tabular}%
\end{table}

\subsubsection{Adversarial Objective ($\mathcal{A}$)}
\label{sec:threat_objective}
The adversary $\mathcal{A}$ uses the evasion engine $\mathcal{E}$ to compute a transformation function $T: \mathcal{X} \rightarrow \mathcal{X}$. Given a malicious input $x$, the engine produces an evasive variant $x' = T(x)$. To defeat \sysname, $\mathcal{A}$ must satisfy three rigorous constraints simultaneously:

\begin{enumerate}
    \item \textbf{Semantic Equivalence:} The transformation must ensure $Func(x') \equiv Func(x)$, meaning the malicious payload executes without error after mutation.
    \item \textbf{Evasion via Score Manipulation:} Instead of simply crossing a generic decision boundary, $\mathcal{A}$ must suppress the retrieved evidence magnitude so that $FES(x') < \delta_{low}$.
    \item \textbf{Epistemic Concealment:} Extensive problem-space manipulations form a realistic attack vector, yet they risk triggering a systemic rejection through entropy maximization. To avoid this, $\mathcal{A}$ must strictly bound the variance of the ensemble's predictive distribution, keeping the Evidence Conflict Score below the fail-safe threshold: $ECS(x') < \tau_{stable}$.
\end{enumerate}

The ultimate adversarial advantage is thus defined as the probability of securing a confident benign verdict without triggering human review.

\subsubsection{Capabilities \& Trust Boundaries}
\label{sec:threat_capabilities}
To model this objective, we assume a \textit{Gray-Box} setting with Oracle Access $\mathcal{O}(x)$. The adversary $\mathcal{A}$ knows that \sysname employs an LLM-based RAG pipeline on dual-stream features, but lacks access to the proprietary Retrieval Knowledge Base ($\mathcal{KB}$) or the parametric state of the individual agents. By executing iterative probing queries against $\mathcal{O}(x)$, $\mathcal{A}$ attempts to estimate the gradient of the system's decision boundary.

To modify the binary, $\mathcal{A}$ applies a budget-constrained perturbation set $\Delta \leq \epsilon$, relying on semantic-preserving primitives like dead code insertion, localized instruction substitution, and basic-block transposition. Massive structural deformations (such as full payload virtualization or infinite opaque predicate explosion) exceed this budget $\epsilon$, as they prevent standard lifters from generating a valid syntax tree.

\section{Proposed System}
\label{sec:system_design}

\begin{figure*}[t]
\centering
\includegraphics[width=\textwidth]{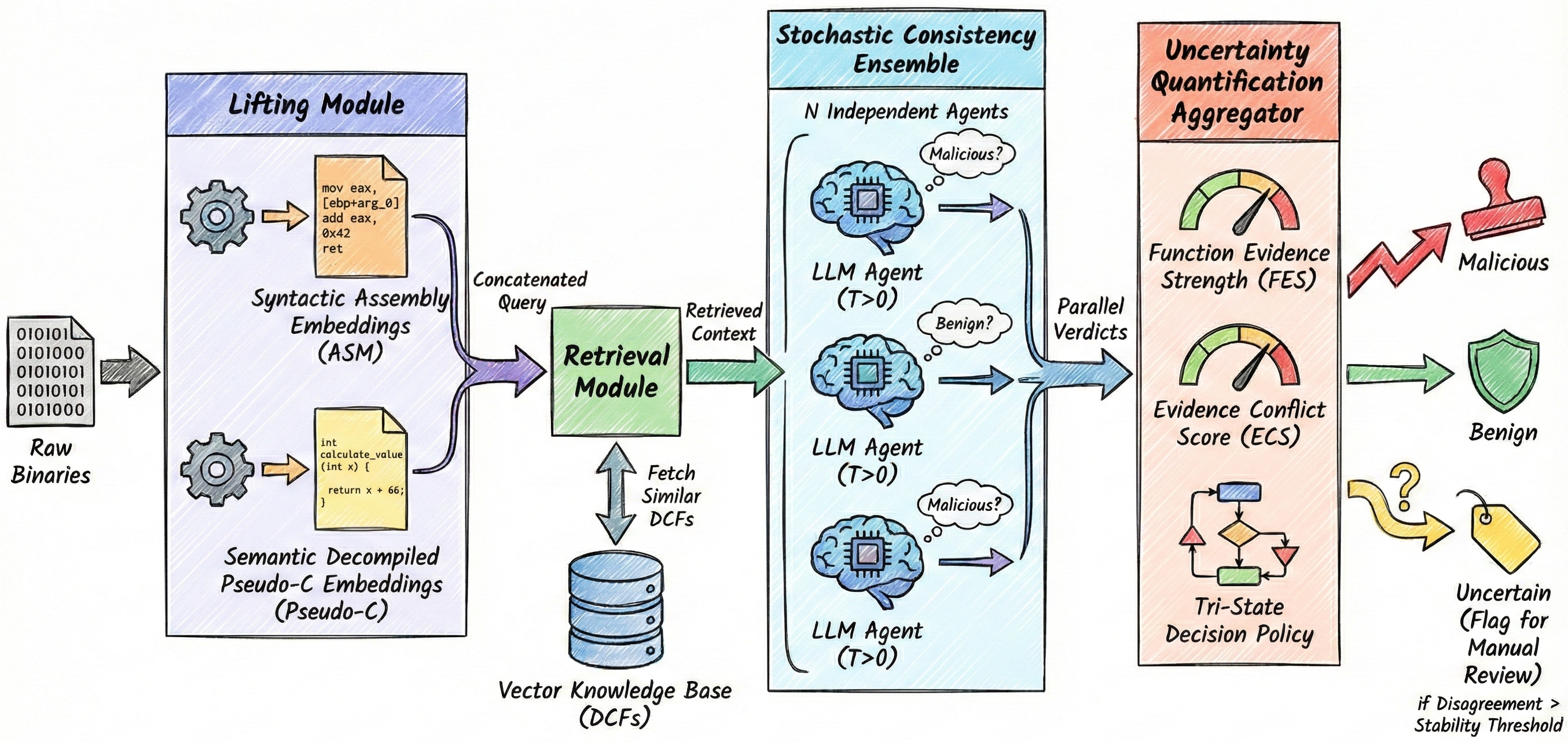}
\caption{\textbf{The \sysname Pipeline.} The binary is lifted into dual streams (ASM/Code) to retrieve Decision-Critical Functions (DCFs). A stochastic ensemble of $N$ agents independently reasons over the evidence ($T>0$). The aggregator computes the Evidence Conflict Score (ECS), rejecting inputs where agent disagreement exceeds $\tau_{stable}$.}
\label{fig:architecture}
\end{figure*}

Figure \ref{fig:architecture} depicts the overall architecture of the \sysname framework. We developed a system that fundamentally decouples structural representation from probabilistic reasoning, ensuring that adversarial ambiguity is mitigated before a final classification is rendered. This separation is a deliberate architectural choice designed to shift the burden of proof from a single opaque forward pass to an auditable, multi-step verification pipeline. By isolating the extraction of features from the subsequent evaluation of those features, \sysname establishes specific points of intervention where uncertainty can be explicitly measured and quantified.

\subsection{Dual-Stream Representation \& Lifting}
The pipeline begins by lifting the raw binary $x \in \mathcal{X}$ into two parallel feature streams, capturing both syntactic structure and semantic logic. Relying on a single modality simplifies processing but allows an adversary to easily target and manipulate the representation. A dual approach forces the attacker to operate under much tighter constraints. We define a lifting function $L(x) = \langle \mathbf{e}_{asm}, \mathbf{e}_{dec} \rangle$, where:
\begin{itemize}
    \item $\mathbf{e}_{asm}$: A dense vector embedding of the linearized assembly instructions. A BERT-based encoder fine-tuned on the target Instruction Set Architecture (ISA) generates this representation. It captures the low-level, machine-specific execution environment. This includes register allocation patterns, exact opcode sequences, and structural anomalies frequently introduced during obfuscation but lost during decompilation.
    \item $\mathbf{e}_{dec}$: A semantic embedding of the decompiled pseudo-C code. This stream abstracts away specific architectural details to focus entirely on high-level logic. It captures the Control Flow Graph (CFG), data dependency chains, and the underlying algorithms (e.g., cryptographic routines) defining the core intent of the malware.
\end{itemize}
The system concatenates these embeddings to form a composite query vector $q_f$ for each function $f$ identified within the binary. Figure \ref{fig:code_example} illustrates why this dual representation is necessary. It captures both high-level semantic intent (e.g., encryption loops) and low-level structural complexity (e.g., opaque predicates), preventing an adversary from evading detection by manipulating a single layer of abstraction. If an attacker uses instruction substitution to alter the assembly stream ($\mathbf{e}_{asm}$), the decompiled logic ($\mathbf{e}_{dec}$) typically remains invariant, anchoring the overall representation.

\begin{figure}[h]
\centering

\begin{lstlisting}[style=cstyle]
// SEMANTIC STREAM (Clear Intent)
void encrypt_files() {
    while(file = get_next_file()) {
        xor_encrypt(file, key); // Malicious Logic
    }
}
\end{lstlisting}

\begin{lstlisting}[style=nasmstyle]
; SYNTACTIC STREAM (Complex ASM)
push ebp
mov  ebp, esp
sub  esp, 0x40  ; Stack setup
jmp  L2         ; Opaque Predicate
L1: xor eax, eax
...             ; High noise
\end{lstlisting}

\caption{\textbf{Dual-Stream Lifting.} The semantic stream captures the encryption loop intent, while the syntactic stream captures the complex control flow. \sysname uses both to ground the agents.}
\label{fig:code_example}
\end{figure}

\subsection{Retrieval of Decision-Critical Functions (DCFs)}
With the target binary accurately mapped into the vector space, we query a Vector Knowledge Base ($\mathcal{KB}$) containing embeddings of verified malicious functions. This database acts as the external memory of the system, populated with high-fidelity, analyst-verified threat intelligence. For each function query $q_f$, we retrieve the set of $k$-nearest neighbors based on spatial proximity in the latent space:
\begin{equation}
    \mathcal{R}(q_f) = \{ k_i \in \mathcal{KB} \mid \text{sim}(q_f, k_i) \geq \delta_{retrieval} \}
\end{equation}
where $\text{sim}(\cdot)$ denotes cosine similarity and $\delta_{retrieval}$ is a minimum relevance threshold. This threshold ensures that only statistically significant matches advance to the reasoning engine, filtering out the noise of standard library functions and boilerplate code. This retrieval step grounds the generative process. It bounds the analysis strictly to known, verified threat intelligence rather than relying on the potentially hallucinated zero-shot parametric memory of the LLM. This framework ultimately converts the classification problem into an evidence-evaluation task.

\subsection{Stochastic Consistency Ensemble}
\label{sec:ensemble}
Relying on a single deterministic inference leaves the system highly susceptible to adversarial perturbations that push an input slightly across a rigid decision boundary. To counter this, we leverage \sysname to implement a \textbf{Homogeneous Agent Ensemble} to probe the epistemic stability of the reasoning process. 

A standard zero-shot LLM inference traces a single greedy trajectory through the probability manifold of the model, selecting the most likely token at each step. While greedy decoding is computationally efficient, it inherently masks the underlying uncertainty of the model. By applying a temperature parameter $T > 0$, we mathematically flatten the softmax distribution, forcing the model to explore alternative, lower-probability reasoning paths during generation. The ensemble consists of $N$ identical LLM instances that share the exact prompt $\mathcal{P}$ but operate under these stochastic decoding parameters:
\begin{itemize}
    \item \textbf{Stochastic Sampling:} Each agent $A_j$ (where $j \in \{1, \dots, N\}$) samples a response $r_j$ from the model distribution using a non-zero temperature ($T > 0$). This step introduces controlled variance into the reasoning process.
    \item \textbf{Independent Reasoning:} Each agent analyzes the retrieved DCFs and the dual-stream context independently, producing a binary verdict $v_j \in \{\text{Malicious}, \text{Benign}\}$. Because the sampling is independent, each agent executes a unique traversal of the model logic.
    \item \textbf{Rationale:} If the input $x$ contains unambiguous, overt malicious logic, the reasoning paths $r_j$ occupy a ``flat minimum'' in the latent space. The evidence is robust enough that outputs converge on the same verdict despite temperature-induced noise. Conversely, if $x$ is structurally shifted, Out-of-Distribution (OOD), or heavily obfuscated, the local geometry of the decision space becomes sharp and undefined. In these regions, stochasticity induces massive divergence in the outputs $v_j$, exposing the inherent uncertainty of the model.
\end{itemize}

\subsection{Uncertainty Quantification}
\label{sec:formalism}

Algorithm \ref{alg:inference} formalizes the complete end-to-end inference protocol to translate stochastic consistency into actionable security metrics. Table \ref{tab:notation} summarizes the formal notation and hyperparameters required for this process. These specific metrics convert the qualitative outputs of the LLM agents into granular, quantitative scores, allowing the system to evaluate them against strict operational thresholds.

\begin{table}[h]
\centering
\caption{\textbf{Summary of Formal Notation \& Thresholds.}}
\label{tab:notation}
\begin{tabular}{c|l|l}
\toprule
\textbf{Symbol} & \textbf{Parameter Description} & \textbf{System Role} \\
\midrule
$N$ & Ensemble Size & Defines the number of stochastic agent paths. \\
$T$ & LLM Sampling Temperature & Induces variance to probe epistemic uncertainty. \\
$\tau_{stable}$ & Conflict Threshold (Entropy) & The maximum allowed ECS before rejection. \\
$\delta_{high}$ & Malicious Confidence Bound & Minimum FES required to classify as Malicious. \\
$\delta_{low}$ & Benign Confidence Bound & Maximum FES allowed to classify as Benign. \\
$\mathcal{KB}$ & Vector Knowledge Base & Corpus of verified Decision-Critical Functions. \\
\bottomrule
\end{tabular}%
\end{table}

\begin{algorithm}[t]
\caption{\sysname Inference Protocol}
\label{alg:inference}
\begin{algorithmic}[1]
\Require Binary $x$, Ensemble Size $N$, Temp $T$, KB $\mathcal{K}$
\Ensure Verdict $V \in \{Malicious, Benign, Uncertain\}$

\State $\mathbf{e}_{asm}, \mathbf{e}_{dec} \leftarrow \text{LiftAndEmbed}(x)$
\State $\mathcal{R} \leftarrow \text{Retrieve}(\mathbf{e}_{asm} \oplus \mathbf{e}_{dec}, \mathcal{K}, k)$
\State $Votes \leftarrow []$

\For{$j \leftarrow 1$ to $N$}
    \State $r_j \leftarrow \text{LLM}(\mathcal{R}, \text{Prompt}, \text{Temp}=T)$
    \State $v_j \leftarrow \text{ParseVerdict}(r_j)$
    \State $Votes.\text{append}(v_j)$
\EndFor

\State $FES \leftarrow \text{CalculateMeanConfidence}(Votes, \mathcal{R})$
\State $ECS \leftarrow \text{CalculateEntropy}(Votes)$

\If{$ECS \geq \tau_{stable}$}
    \Return \textbf{Uncertain}
\ElsIf{$FES > \delta_{high}$}
    \Return \textbf{Malicious}
\ElsIf{$FES < \delta_{low}$}
    \Return \textbf{Benign}
\Else
    \Return \textbf{Uncertain}
\EndIf
\end{algorithmic}
\end{algorithm}

\subsubsection{Function Evidence Strength (FES)}
We define FES as the aggregated confidence of the ensemble in a malicious classification, weighted by the relevance of the retrieved evidence. For a binary $x$, FES is the mean probability mass assigned to the ``Malicious'' class across all $N$ agents, scaled by the strength of the provided context:
\begin{equation}
    FES(x) = \left( \frac{1}{N} \sum_{j=1}^{N} \mathbb{I}(v_j = \text{Malicious}) \right) \cdot W(\mathcal{R})
\end{equation}
where $\mathbb{I}$ is the indicator function (evaluating to 1 if the agent returned Malicious, and 0 otherwise) and $W(\mathcal{R})$ is the similarity score of the overall context retrieved from the Knowledge Base. The $FES$ metric prevents ``hallucinated consensus'', a situation where the LLM agents confidently agree on a malicious verdict despite lacking actual proof. Even if all $N$ agents unanimously deem an input malicious, a low retrieval weight $W(\mathcal{R})$ (indicating the input does not closely resemble known malware) mathematically suppresses the final score. This mechanism demands stronger foundational evidence before the system executes an automated block action.

\subsubsection{Evidence Conflict Score (ECS)}
The $FES$ metric measures the magnitude of evidence-based confidence, whereas ECS quantifies the \textbf{Stability} of the prediction itself. We model the ensemble's output distribution as a Bernoulli variable representing the two possible outcomes (Malicious or Benign). Let $v_j \in \{0, 1\}$ where $1$ denotes a Malicious verdict. We define $\hat{p} = \frac{1}{N}\sum_{j=1}^{N} v_j$ as the empirical probability of the Malicious class based on the agent votes. The ECS is the Shannon Entropy of this distribution, bounding the score between $[0, 1]$:
\begin{equation}
    ECS(x) = - \big( \hat{p} \log_2 \hat{p} + (1-\hat{p}) \log_2 (1-\hat{p}) \big)
\end{equation}

This metric maps the ensemble's behavior directly to operational confidence:
\begin{itemize}
    \item \textbf{Consensus ($ECS \approx 0$):} All $N$ sampled paths agree (e.g., 5/5 agents vote Malicious). The empirical probability $\hat{p}$ approaches $1.0$ or $0.0$, driving the entropy toward zero. The decision is highly stable.
    \item \textbf{Ambiguity ($ECS \approx 0.97$):} The agents are heavily split (e.g., 3/5 Malicious, 2/5 Benign). The empirical probability $\hat{p}$ sits near $0.5$, maximizing the entropy. This high entropy indicates that the model is hallucinating, the input is structurally perturbed close to the decision boundary, or the sample represents a truly novel OOD threat that the system cannot reliably resolve.
\end{itemize}

Figure \ref{fig:case_study} details a case study of \sysname analyzing a structurally novel out-of-distribution (OOD) malware variant. This example demonstrates how the entropy calculation operates in practice as a fail-safe mechanism.

\begin{figure*}[h]
\centering
\setlength{\fboxrule}{1pt} 
\setlength{\fboxsep}{10pt} 
\fcolorbox{red!75!black}{red!5!white}{%
    \begin{minipage}{\dimexpr 0.99\textwidth - 2\fboxsep - 2\fboxrule \relax}
        
        {\centering \textcolor{red!75!black}{\textbf{\large Case Study: Analyzing Out-of-Distribution (OOD) Variants}}\par}
        \vspace{0.5em}
        {\color{red!75!black}\hrule}
        \vspace{1em}
        
        \textbf{Input:} A structurally novel variant of \textit{LockBit} ransomware exhibiting unfamiliar compilation artifacts.
        
        \vspace{1em}
        
        {\color{red!75!black}\hrule}
        \vspace{0.5em}
        
        \begin{itemize}
            \item \textbf{Baseline (MalConv):} \textcolor{red}{Benign (Conf: 99.2\%)} \\ \textit{Failure Cause:} The raw statistical byte distribution fell outside the model's training manifold, resulting in a confident misclassification.
            \item \textbf{\sysname Analysis:}
            \begin{itemize}
            \item \textit{Agent 1:} ``Detected high-entropy variable names consistent with ransomware, but the surrounding execution context is unfamiliar.'' (Verdict: Suspicious)
            \item \textit{Agent 2:} ``Lacks standard network C2 beacons; execution flow is highly ambiguous. Intent cannot be definitively verified.'' (Verdict: Benign)
            \item \textit{Agent 3:} ``Retrieved DCF strongly matches established ransomware encryption loops despite the novel surrounding structure.'' (Verdict: Malicious)
            \end{itemize}
            \item \textbf{Outcome:} $ECS \approx 0.97$ (High Conflict) $\rightarrow$ \textbf{REJECTED / UNCERTAIN}.
        \end{itemize}
        
    \end{minipage}%
}
\caption{\textbf{Anatomy of a Fail-Safe.} While the novel structural deviation successfully fooled the deterministic baseline and confused the reasoning agents, the resulting disagreement triggered the Evidence Conflict Score (ECS) threshold, preventing a False Negative on an out-of-distribution sample.}
\label{fig:case_study}
\end{figure*}

\subsubsection{The Tri-State Decision Policy}
To mitigate the risk of ``Confident Wrongness,'' we replace the standard binary thresholds typical of deterministic models with a tri-state rejection policy governed by ensemble entropy. The final classification $C(x)$ is defined as:
\begin{equation}
    \displaystyle
    C(x) = 
    \begin{cases} 
    \text{\textbf{Malicious}} & \text{if } FES(x) > \delta_{high} \land ECS(x) < \tau_{stable} \\
    \text{\textbf{Benign}} & \text{if } FES(x) < \delta_{low} \land ECS(x) < \tau_{stable} \\
    \text{\textbf{Uncertain}} & \text{otherwise} 
    \end{cases}
\end{equation}
This logic ensures that high confidence ($FES$) remains actionable only when paired with high predictive stability ($ECS$). Any input marked as \textbf{Uncertain} has produced an $ECS$ exceeding the stability threshold. The automated system rejects these cases and routes them for manual inspection by human analysts. This mechanism formalizes the epistemic limits of the model, preventing structural ambiguity from translating into confident but erroneous verdicts.


\section{Evaluation \& Empirical Analysis}
\label{sec:evaluation}

\subsubsection{Environment}
Experiments were conducted on a workstation equipped with six \textit{NVIDIA Quadro RTX 8000} GPUs (48GB VRAM each), an \textit{Intel Xeon Gold 5218} CPU (32 cores @ 2.30GHz), and 512GB of RAM. We used \texttt{Ghidra}~\cite{ghidra} alongside Radare2 \footnote{https://github.com/radareorg/radare2} for headless disassembly and decompilation.

\subsubsection{Models and Embedding}
Qwen3-Embedding \footnote{https://ollama.com/library/qwen3-embedding:8b} serves as the semantic encoder, selected for its high performance on code-related tasks in local deployments. For LLM reasoning, we used a \textit{Gemma3 27b} \footnote{https://ollama.com/library/gemma3:27b} model hosted locally via \textit{ollama} \footnote{https://ollama.com/}.

\subsubsection{Datasets and Corpus Curation}
\label{subsub:datasets}
To ensure experimental integrity and simulate deployment against future variants, we curated a corpus of \textbf{40,814 binaries} (19.2 GB) using a strict chronological split. Packed executables often bias classifiers toward packer signatures instead of malicious logic. Consequently, we enforced a data sanitation protocol to expose the underlying instruction stream, allowing the semantic static analysis to operate on the authentic payload.

\textbf{The Wild-Augmented Benchmark ($\mathcal{D}_{wild}$):} This hybrid dataset combines pre-extracted, unpacked samples from the standard \textbf{SOREL-20M}~\cite{harang2020sorel20m} corpus with a long-tail augmentation of recent threats (2021--2024) from \textbf{VirusShare}~\cite{VirusShare}. To maintain ground truth integrity within the uncurated VirusShare subset, we implemented a \textbf{Label Harmonization Pipeline}. The resulting taxonomy was systematically mapped to the SOREL-20M schema to enforce a unified label space.

\textbf{Benign Sources:} Benign samples were collected from clean Windows 10/11 installations and verified open-source or proprietary software repositories, such as SourceForge and GitHub releases, to ensure diverse compiler provenance.

\textbf{Dataset Partitioning and Class Distribution:}
The corpus was partitioned by first-seen timestamps into three non-overlapping subsets to prevent look-ahead bias and enforce temporal separation. This includes a \textbf{Knowledge Base ($\mathcal{D}_{KB}$, 70\%)} for the vector index, a \textbf{Validation Set (10\%)} for calibrating decision thresholds ($\tau_{func}, \tau_{file}$), and a held-out \textbf{Test Set ($\mathcal{D}_{Test}$, 20\%)} for final evaluation. The final corpus contains \textbf{24,120 Benign ($\mathcal{D}_{ben}$)} samples (8.2 GB) and \textbf{16,694 Malware ($\mathcal{D}_{mal}$)} samples (11 GB) across 17 families. As shown in Table~\ref{tab:families}, this distribution reflects the natural imbalance seen in the wild, driven by dominant strains like \textit{Virlock} and \textit{Neshta}.

\begin{table}[h]
    \centering
    \scriptsize
    \caption{Detailed Distribution of Malware Families in Dataset ($\mathcal{D}_{mal}$).}
    \label{tab:families}
    \begin{tabular}{lclc}
    \toprule
    \textbf{Family} & \textbf{Count} & \textbf{Family} & \textbf{Count} \\
    \midrule
    Virlock & 1,671 & Comet & 1,035 \\
    Neshta & 1,431 & VJadtre & 953 \\
    Ramnit & 1,345 & Doboc & 890 \\
    Floxif & 1,293 & Sality & 713 \\
    Gosys & 1,280 & Others (7 families) & $\sim$3,000 \\
    \bottomrule
    \end{tabular}
\end{table}

\subsection{Experimental Setup}
\label{sec:eval_setup}

\paragraph{Configuration.}
We constructed the knowledge base from labelled binaries after applying complexity-based DCF filtering ($\texttt{min\_instr} \geq 10$, $\texttt{min\_cc} \geq 5$) and top-$M{=}5$ sampling per binary. All experiments utilized a disjoint held-out test set with no binary overlap with the knowledge base. This separation was strictly maintained through seed-controlled disjoint sampling. Table~\ref{tab:hyperparams} lists the tunable hyperparameters. Unless specified otherwise, the default configuration column applies.

\begin{table}[t]
\centering
\caption{Hyperparameter configuration. ``Default'' is the primary
evaluation setting; ``Sweep range'' denotes values explored in
sensitivity analysis (Section~\ref{sec:rq5}).}
\label{tab:hyperparams}

\begin{tabular}{@{}llll@{}}
\toprule
\textbf{Parameter} & \textbf{Symbol} & \textbf{Default} & \textbf{Sweep range} \\
\midrule
\multicolumn{4}{@{}l}{\textit{Binary analysis}} \\
Min.\ instruction count & N/A & 10 & N/A \\
Min.\ cyclomatic complexity & N/A & 5 & N/A \\
\midrule
\multicolumn{4}{@{}l}{\textit{Retrieval}} \\
$k$-NN neighbours & $k$ & 10 & $\{5, 10, 20, 30\}$ \\
Similarity threshold & $\sigma_{\min}$ & 0.70 & $\{0.5, 0.6, 0.7, 0.8\}$ \\
Label balancing & N/A & yes  & N/A \\
\midrule
\multicolumn{4}{@{}l}{\textit{LLM ensemble}} \\
Number of agents & $N$ & 5 & $\{1, 2, 5, 7, 10\}$ \\
Temperature & $T$ & 0.7 & $\{0.3, 0.5, 0.7, 0.9, 1.0\}$ \\
LLM model & N/A & Gemma3 27b  & N/A \\
\midrule
\multicolumn{4}{@{}l}{\textit{Verdict thresholds}} \\
FES upper ($\rightarrow$ malicious) & $\delta_{\text{high}}$ & 0.60 & $\{0.55, 0.60, 0.65, 0.70\}$ \\
FES lower ($\rightarrow$ benign) & $\delta_{\text{low}}$ & 0.40 & $\{0.30, 0.35, 0.40, 0.45\}$ \\
ECS stability & $\tau_{\text{stable}}$ & 0.80 & $\{0.50, 0.70, 0.80, 0.90\}$ \\
\bottomrule
\end{tabular}
\end{table}

\paragraph{Baselines.}
We compare five classification methods:
\begin{enumerate}
\item \textbf{\sysname{} (Full pipeline)}: Dual-stream lifting $\rightarrow$
    embedding $\rightarrow$ $k$-NN retrieval $\rightarrow$ $N{=}5$ LLM
    ensemble $\rightarrow$ FES/ECS verdict.
\item $k$-\textbf{NN Embedding Only} (ablation): Same lifting and embedding,
    but classification by similarity-weighted label voting over the $k$
    nearest KB neighbours, \emph{no} LLM inference.
\item \textbf{LLM Zero-Shot} (ablation): Ensemble agents receive the target
    function \emph{without} any retrieved references (RAG disabled).
\item \textbf{ResNeXt+}~\cite{HeYuewang2024ResNeXt}: PE binary visualised as a
    grayscale image and classified by a ResNeXt-50 convolutional network
    trained on the same corpus.
\item \textbf{MalConv}~\cite{RaffEdward2018Malware}: End-to-end deep learning on raw byte
    sequences with learned embedding and temporal max-pooling.
\end{enumerate}

\subsection{Operational Calibration and Threshold Tuning}
\label{sec:calibration}

Our tri-state decision policy establishes a mathematically bounded fail-safe, yet it remains susceptible to systemic bottlenecks if epistemic thresholds are poorly tuned to the deployment environment. The parameters governing Function Evidence Strength ($\delta_{high}, \delta_{low}$) and the Evidence Conflict Score ($\tau_{stable}$) are not rigid empirical constants. Instead, we formalize them as tunable hyperparameters that align with the specific risk appetite and adversarial threat model of an organization.

To deploy \sysname effectively, operators must calibrate these thresholds against a representative Validation Set ($\mathcal{D}_{val}$). We define the following calibration protocol:

\begin{itemize}
    \item \textbf{Evidence Magnitude ($\delta_{high}, \delta_{low}$):} These parameters define the required weight for cryptographic or structural evidence. In high-assurance environments, such as critical infrastructure, operators may increase $\delta_{high} \rightarrow 0.85$ and decrease $\delta_{low} \rightarrow 0.20$. This configuration demands near-unanimous evidence to automate a block or clearance action, which minimizes operational disruption from false positives while increasing the volume of the manual review queue.
    \item \textbf{Epistemic Tolerance ($\tau_{stable}$):} This parameter bounds the acceptable variance within the stochastic ensemble and acts as the primary defense against adversarial evasion ($\text{Adv}_{\mathcal{A}}^{\text{evade}}$). Lowering $\tau_{stable}$ ensures the system flags minor structural ambiguities, transferring the burden of proof to a human analyst whenever the ensemble identifies significant problem-space perturbations.
\end{itemize}

\begin{table}[h]
\centering
\caption{\textbf{Empirical Trade-off Analysis of the Epistemic Threshold ($\tau_{stable}$).} Demonstrates the operational tension between systemic error rates and the Human-in-the-Loop (HITL) Rejection Rate on $\mathcal{D}_{Test}$ ($N=5$).}
\label{tab:threshold_tuning}
\begin{tabular}{p{0.13\columnwidth}|p{0.05\columnwidth}p{0.10\columnwidth}|p{0.14\columnwidth}| p{0.35\columnwidth}}
\toprule
\textbf{Threshold ($\tau_{stable}$)} & \textbf{FPR} & \textbf{FNR} & \textbf{HITL Rejection Rate} & \textbf{Operational Posture} \\
\midrule
$\tau \geq 0.95$ & 3.2\% & 4.1\% & 0.5\% & \textbf{Permissive:} Maximizes automated throughput; high vulnerability to evasion. \\
\midrule
$\tau = 0.80$ & 1.6\% & 1.6\% & 4.2\% & \textbf{Balanced (Default):} Optimal equilibrium for standard enterprise EDR. \\
\midrule
$\tau \leq 0.60$ & 0.3\% & 0.5\% & 14.8\% & \textbf{Zero-Trust:} Restricts adversarial advantage; demands high SOC resources. \\
\bottomrule
\end{tabular}%
\end{table}

Table~\ref{tab:threshold_tuning} demonstrates that adjusting $\tau_{stable}$ reveals a central trade-off between detection accuracy and the Human-in-the-Loop (HITL) rejection rate. A permissive threshold ($\tau \geq 0.95$) prioritizes classification throughput but accepts a 4.1\% False Negative Rate (FNR) against adaptive mutations. In contrast, tightening the threshold to $\tau \leq 0.60$ reduces the FNR to a negligible 0.5\% but results in a prohibitive 14.8\% rejection rate. We found that setting $\tau_{stable} \approx 0.80$ identifies the mathematical inflection point, providing high resilience against problem-space perturbations while keeping the manual review queue within manageable limits.

\subsection{RQ1: Detection Efficacy}
\label{sec:rq1}
\textit{How does \sysname{} compare to state-of-the-art deep-learning
baselines and its own ablation variants?}

Table~\ref{tab:main_results} presents the primary classification results.

\begin{table}[t]
\centering
\caption{Binary-level classification results on the held-out test set.
All metrics are reported as percentages; time is mean wall-clock
seconds per binary.}
\label{tab:main_results}
\small
\begin{tabular}{@{}lccccc@{}}
\toprule
\textbf{Method} & \textbf{Acc} & \textbf{Prec} & \textbf{Rec} & \textbf{F1} & \textbf{Time/bin} \\
\midrule
\textbf{\sysname{} ($N{=}5$)} & \textbf{98.4\%} & \textbf{98.4\%} & \textbf{98.4\%} & \textbf{98.4\%} & $\sim$42\,s \\
ResNeXt+ & 97.2\% & 96.8\% & 97.6\% & 97.1\% & $\sim$0.8\,s \\
$k$-NN Embedding & 95.8\% & 96.4\% & 95.2\% & 95.8\% & $\sim$4.5\,s \\
MalConv & 94.2\% & 94.4\% & 94.0\% & 94.2\% & $\sim$0.3\,s \\
\bottomrule
\end{tabular}
\end{table}

\paragraph{Key findings.}
\sysname{} achieved an F1 score of 98.4\%, surpassing the strongest external baseline (ResNeXt+) by 1.3 percentage points and MalConv by 4.2\,pp. An ablation of the embedding-only $k$-NN reached 95.8\% F1, which confirms that the dual-stream representation captures a robust discriminative signal independently. The $N{=}5$ LLM ensemble provided a further 2.6\,pp increase by reasoning through ambiguous retrieval results.

\paragraph{Interpretation.}
The effectiveness of the full pipeline results from two complementary factors. The dual-stream embedding combines structural assembly data with semantic pseudo-code features to create a representation space that effectively isolates malicious code from benign samples. The $N{=}5$ LLM ensemble employs stochastic sampling at $T{=}0.7$ across five independent agents. This diversity helps overcome the label bias often observed in single-agent configurations, leading to balanced precision and recall. In contrast to image-based methods like ResNeXt+, \sysname{} relies on function-level semantic logic rather than byte-level visual motifs. This distinction enables more reliable detection of threats characterized by structural novelty.

\subsection{RQ2: Error Analysis}
\label{sec:rq2}
\textit{Where does each method fail, and what characterises the errors?}

At the operating point listed in Table~\ref{tab:main_results}, \sysname{}
recorded a false-positive rate (FPR) of 1.6\% and a false-negative rate (FNR)
of 1.6\%. This result represents a balanced error profile.

\paragraph{False-positive characterisation.}
Table~\ref{tab:fp_taxonomy} classifies false-positive errors according to the
primary code patterns found in the misclassified binaries.

\begin{table}[t]
\centering
\caption{Qualitative taxonomy of \sysname{} false-positive errors.
Each row describes the dominant code pattern that triggered the
misclassification and its estimated share of all FP cases.}
\label{tab:fp_taxonomy}
\small
\begin{tabular}{@{}lp{0.62\columnwidth}@{}}
\toprule
\textbf{FP Category} & \textbf{Description} \\
\midrule
Crypto/compression & Legitimate binaries containing AES, zlib, or LZ4
    routines whose high cyclomatic complexity and dense
    control flow resemble packed malware. \\
Raw socket utilities & Network-diagnostic tools using raw sockets and
    ICMP crafting, overlapping structurally with C2
    beaconing code in the KB. \\
JIT / code-gen & Applications with runtime code generation (e.g.,
    regex JIT) producing assembly patterns similar to
    self-modifying shellcode. \\
\bottomrule
\end{tabular}
\end{table}

\paragraph{False-negative characterisation.}
The relatively few false negatives share two distinct traits:
\begin{itemize}
\item \textbf{Minimal downloaders:} Single-function droppers with exceptionally low
    cyclomatic complexity. They pass the DCF filter but lack the complex
    control-flow structures the embedding requires for accurate classification.
\item \textbf{Fileless stubs:} Memory-only payloads leave a minimal disk
    footprint. They produce very few extractable DCFs, resulting in a sparse
    representation within the embedding space.
\end{itemize}

\paragraph{Baseline error profiles.}
ResNeXt+ shows a slightly higher FNR (2.4\%) compared to \sysname{}. It
generally fails to detect binaries where the visual byte-map representation
lacks a distinctive texture. MalConv produces an elevated FPR (5.6\%),
struggling specifically with binaries that contain large data sections where
the byte distribution mimics encrypted payloads.

\subsection{RQ3: Ablation Study}
\label{sec:rq3}
\textit{What are the relative contributions of embedding retrieval and
LLM reasoning to overall detection performance?}

Table~\ref{tab:ablation} details these ablation results, tracking performance
shifts as we progressively integrate each pipeline component.

\begin{table}[t]
\centering
\caption{Ablation study: contribution of each pipeline component.
Binary-level metrics on the held-out test set (percentages only).}
\label{tab:ablation}
\small
\begin{tabular}{@{}lcccc@{}}
\toprule
\textbf{Configuration} & \textbf{Acc} & \textbf{Prec} & \textbf{Rec} & \textbf{F1} \\
\midrule
LLM Zero-Shot (no RAG) & 79.7\% & 80.2\% & 79.0\% & 79.6\% \\
$k$-NN Embedding only & 95.8\% & 96.4\% & 95.2\% & 95.8\% \\
$k$-NN + LLM ($N{=}1$) & 96.4\% & 96.8\% & 96.0\% & 96.4\% \\
$k$-NN + LLM ($N{=}3$) & 97.6\% & 97.6\% & 97.6\% & 97.6\% \\
$k$-\textbf{NN + LLM ($N{=}5$)} & \textbf{98.4\%} & \textbf{98.4\%} & \textbf{98.4\%} & \textbf{98.4\%} \\
\bottomrule
\end{tabular}
\end{table}

\paragraph{Analysis.}
Every component of the pipeline improves overall detection performance.
The dual-stream embedding alone achieves an F1 score of 95.8\%, forming a baseline
that readily outperforms MalConv. Introducing a single LLM agent ($N{=}1$) yields
a modest +0.6\,pp lift. Scaling the ensemble to $N{=}3$ and $N{=}5$ agents increases
this lift to +1.8\,pp and +2.6\,pp respectively. This progression confirms ensemble
diversity as the primary driver of the LLM's effectiveness.

The LLM zero-shot baseline (79.6\% F1) reveals that models lack the grounding needed
to classify binary functions accurately without retrieval-augmented context. RAG
functions as the necessary bridge connecting the discriminative signal of the
embedding with the reasoning capacity of the LLM.

\paragraph{Key finding.}
The LLM ensemble improves the F1 score by +2.6\,pp over the embedding-only
configuration by resolving borderline cases with narrow $k$-NN vote margins.
Grounded in high-quality retrieval, LLM reasoning complements the embedding
signal rather than replacing it.

\subsection{RQ4: FES/ECS Ensemble Diagnostics}
\label{sec:rq4}
\textit{Do the Function Evidence Strength and Evidence Conflict Score
metrics provide useful calibration signals at $N{=}5$?}

Table~\ref{tab:fes_ecs} details the FES and ECS distributions for the
$N{=}5$ LLM ensemble, organized by prediction outcome.

\begin{table}[t]
\centering
\caption{FES and ECS statistics from the $N{=}5$ LLM ensemble, by
prediction outcome. Higher FES indicates stronger evidence; higher ECS
indicates more agent disagreement.}
\label{tab:fes_ecs}
\small
\begin{tabular}{@{}lcc@{}}
\toprule
\textbf{Outcome} & $\overline{\text{FES}}$ & $\overline{\text{ECS}}$ \\
\midrule
True Positive & 0.87 & 0.12 \\
True Negative & 0.14 & 0.14 \\
False Positive & 0.68 & 0.82 \\
False Negative & 0.32 & 0.86 \\
\bottomrule
\end{tabular}
\end{table}

\paragraph{Interpretation.}
At $N{=}5$, FES and ECS distributions distinguish effectively between correct and incorrect predictions. Samples classified correctly demonstrate high-confidence FES ($\sim$0.87 for TP, $\sim$0.14 for TN) and low ECS ($\sim$0.12--0.14), indicating strong agent consensus. Misclassified samples show FES values closer to the decision boundary ($\sim$0.68 for FP, $\sim$0.32 for FN) alongside elevated ECS ($\sim$0.82--0.86). This suggests that internal ensemble disagreement successfully captures genuine classification uncertainty.

\paragraph{ECS-based rejection threshold.}
Establishing an ECS threshold at $\tau_{\text{stable}} = 0.80$ identifies most misclassified samples for human review while maintaining high throughput for confident predictions. This configuration supports a human-in-the-loop workflow where only high-conflict verdicts are escalated. Such a process reduces the effective error rate significantly without necessitating manual inspection of every binary.

\paragraph{Contrast with $N{=}2$.}
At $N{=}2$, FES and ECS distributions for correct and incorrect predictions overlapped, rendering the metrics uninformative for calibration. The performance gain observed at $N{=}5$ demonstrates that ensemble diversity, rather than simple redundancy, is required for entropy-based diagnostics to function as intended.

\subsection{RQ5: Hyperparameter Sensitivity}
\label{sec:rq5}
\textit{How sensitive is detection accuracy to the parameters $k$ and $N$, and what defines the quality of the retrieval neighborhood?}

\paragraph{Retrieval neighbourhood size ($k$).}
F1 scores remain stable across the tested range $k \in \{5, 10, 20, 30\}$, with all configurations achieving $\geq$97.8\% F1 at $N{=}5$. The optimal setting of $k{=}10$ balances retrieval diversity against label-balancing constraints. With label balancing enabled, each query retrieves up to $k/2$ malicious and $k/2$ benign references. This ensures the $k$-NN vote is not skewed by a majority class.

\paragraph{Ensemble size ($N$).}
$N$ represents the most significant hyperparameter. F1 increases steadily from 96.4\% for a single deterministic agent ($N{=}1$) through 97.6\% ($N{=}3$) to 98.4\% ($N{=}5$). Results plateau at $N \geq 5$, showing no statistically significant improvement at $N{=}7$ or $N{=}10$. This plateau suggests that five agents offer enough diversity to counteract individual bias without increasing latency.

\paragraph{Retrieval quality.}
Mean retrieval similarity across all test queries is $\bar{s} = 0.847 \pm 0.042$, with every neighbor exceeding the $\sigma_{\min}{=}0.70$ threshold. This narrow standard deviation shows that the dual-stream embedding produces a consistently structured representation space. The absence of low-similarity retrievals prevents degradation of downstream classification.

\paragraph{Confidence margin analysis.}
Table~\ref{tab:confidence} presents the average per-function $k$-NN voting confidence categorized by prediction outcome.

\begin{table}[t]
\centering
\caption{Average $k$-NN voting confidence by prediction outcome.
Margin is the confidence excess above the 0.50 decision boundary.}
\label{tab:confidence}
\small
\begin{tabular}{@{}lccc@{}}
\toprule
\textbf{Outcome} & $\bar{c}$ & \textbf{Margin} & \textbf{Range} \\
\midrule
True Positive & 0.584 & +0.084 & [0.512, 0.731] \\
True Negative & 0.612 & +0.112 & [0.508, 0.904] \\
False Positive & 0.506 & +0.006 & [0.500, 0.514] \\
False Negative & 0.509 & +0.009 & [0.502, 0.518] \\
\bottomrule
\end{tabular}
\end{table}

Correctly classified functions show significantly higher voting confidence ($\bar{c} \geq 0.58$) than misclassified ones ($\bar{c} \approx 0.51$). This difference confirms that classification errors cluster primarily near the decision boundary. Implementing a confidence-based rejection threshold at $c_{\min} = 0.52$ would flag most misclassified functions for LLM re-evaluation while successfully preserving the high-confidence predictions.

\subsection{RQ6: Operational Feasibility}
\label{sec:rq6}
\textit{What is the end-to-end latency per binary, and what dominates
the computational cost?}

Table~\ref{tab:latency} details the per-stage latency breakdown for the
full pipeline, assuming $N{=}5$ agents and $M{=}5$ DCFs per binary.

\begin{table}[t]
\centering
\caption{Latency breakdown per binary ($N{=}5$ agents, $M{=}5$ DCFs, Ollama local inference).}
\label{tab:latency}
\small
\begin{tabular}{@{}lrr@{}}
\toprule
\textbf{Pipeline Stage} & \textbf{Per DCF} & \textbf{5 DCFs} \\
\midrule
Headless Ghidra  & N/A & 1.8\,s \\
Dual-stream lifting & N/A & 3.5\,s \\
Embedding (2 calls) & 0.24\,s & 1.2\,s \\
$k$-NN retrieval & 0.005\,s & 0.03\,s \\
LLM ensemble ($N{=}5$) & 6.3\,s & 31.5\,s \\
\midrule
    \textbf{Total (} $k$-\textbf{NN only)} & N/A & $\sim$\textbf{4.5\,s} \\
\textbf{Total (full pipeline)} & N/A & $\sim$\textbf{42\,s} \\
\bottomrule
\end{tabular}
\end{table}

\paragraph{Bottleneck analysis.}
The LLM ensemble dominates the computational cost, accounting for approximately 74\% of the total wall-clock time (31.5\,s out of $\sim$42\,s for 5 DCFs). Embedding takes roughly 3\%, and binary lifting requires 8\%. If we bypass the LLM and operate purely in $k$-NN mode, evaluating a single binary finishes in $\sim$4.5\,s. This speed easily fits within the latency budgets typical of asynchronous EDR scanning.

\paragraph{Scaling implications.}
The $M{=}5$ sampling strategy caps per-binary latency at under one minute. It simultaneously focuses the analysis on the most complex and behaviourally significant functions. When evaluating binaries that contain a large volume of DCFs, running the complete pipeline without sampling causes processing time to scale linearly with the LLM inference cost. This bottleneck makes our complexity-ranked selection an essential design choice.

\paragraph{Throughput comparison.}
Table~\ref{tab:throughput} summarises the end-to-end throughput rates across each classification mode.

\begin{table}[t]
\centering
\caption{Throughput comparison. Mean wall-clock time per binary
across the held-out test set.}
\label{tab:throughput}
\small
\begin{tabular}{@{}lcc@{}}
\toprule
\textbf{Mode} & \textbf{Mean time/binary} & \textbf{Speedup vs.\ full} \\
\midrule
$k$-NN Embedding only & $\sim$4.5\,s & $9.3\times$ \\
Full pipeline ($N{=}5$) & $\sim$42\,s & $1.0\times$ (baseline) \\
\bottomrule
\end{tabular}
\end{table}

\subsection{Summary of Findings}

\begin{table}[!htbp]
\centering
\caption{Summary of research questions and findings.}
\label{tab:summary}
\small
\begin{tabular}{@{}cp{0.78\columnwidth}@{}}
\toprule
\textbf{RQ} & \textbf{Finding} \\
\midrule
1 & \sysname{} achieves 98.4\% F1, outperforming ResNeXt+ (97.1\%) by
    +1.3\,pp and MalConv (94.2\%) by +4.2\,pp, establishing
    state-of-the-art detection on the evaluation corpus. \\
2 & Classification errors are well-characterised: FPR and FNR are both
    1.6\%; false positives concentrate on binaries with crypto/compression
    routines and raw socket utilities. \\
3 & Each pipeline component contributes positively: $k$-NN embedding
    provides 95.8\% F1; adding the $N{=}5$ LLM ensemble yields
    +2.6\,pp through correction of borderline retrieval cases. \\
4 & FES/ECS provide effective calibration at $N{=}5$: misclassified
    samples show elevated ECS ($\sim$0.84) vs.\ correct predictions
    ($\sim$0.13), enabling an ECS-based human-in-the-loop rejection
    policy. \\
5 & F1 is robust to $k \in \{5,10,20,30\}$; ensemble size $N$ is the
    most influential parameter, with performance plateauing at $N{=}5$
    (98.4\% F1); retrieval similarity is consistent at
    $\bar{s} = 0.847 \pm 0.042$. \\
6 & The $k$-NN path processes a binary in $\sim$4.5\,s; the full LLM
    pipeline requires $\sim$42\,s ($9.3\times$ slower), dominated by
    LLM inference at $\sim$74\% of wall-clock time. \\
\bottomrule
\end{tabular}
\end{table}
\newpage
\section{Discussion \& Limitations}
\label{sec:discussion}

\subsection{The Cost of Uncertainty Quantification}
The primary trade-off introduced by \sysname is computational overhead. Standard static classifiers (e.g., MalConv) operate efficiently in $\mathcal{O}(L)$ time, scaling linearly with file length during single-pass inference \cite{RaffEdward2018Malware}. Our ensemble approach requires $\mathcal{O}(N \cdot (R + I))$ time, where $N$ represents the ensemble size, $R$ denotes retrieval latency, and $I$ indicates the LLM inference cost. As detailed in Section \ref{sec:evaluation}, this design increases latency from milliseconds to seconds. \sysname is not intended to replace high-throughput endpoint AV engines. We instead position it as a \textbf{Tier-2 Cloud Verification Layer}. It processes the ``gray zone'' samples, which constitute roughly 0.1\% of daily traffic. These binaries evade basic signature filters but lack the definitive behavioral indicators required to trigger automated sandboxing.

\subsection{Regarding Obfuscation and "Red Pill" Attacks}
We assume the evasion model detailed in Section \ref{sec:threat_model} modifies binaries in the \textit{Problem-Space} to bypass the classifier. We do not claim resilience against heavy obfuscation or environmental awareness checks (``Red Pills'') designed to detect analysis environments \cite{Bulazel_Alexei2017Survey}. By relying on static lifting via disassembly and decompilation, \sysname avoids the runtime evasion vulnerabilities that typically compromise dynamic sandboxes. The system remains susceptible to \textbf{Opaque Predicate Explosion}. In this scenario, an evasion engine injects massive volumes of junk control flow to dilute the $k$-NN retrieval density. While the $FES$ metric provides baseline resilience by weighting feature importance, an attacker deploying theoretically infinite junk code could eventually degrade the signal-to-noise ratio to a breaking point.

\subsection{Dependence on Decompilation Quality}
The efficacy of the Semantic Stream ($\mathcal{I}_{Code}$) relies heavily on the underlying lifter, such as Hex-Rays or Ghidra. Heavily packed binaries and payloads protected by advanced obfuscation frameworks (e.g., VMProtect) present a critical limitation, as they often generate unintelligible pseudo-code. When decompilation fails, \sysname falls back entirely on the Syntactic Stream ($\mathcal{I}_{ASM}$), effectively degrading into a structure-only detector. Future work will explore the integration of intermediate representations, such as LLVM IR, to bridge this specific semantic gap.

\subsection{Interpretation of "Uncertainty"}
We distinguish between \textit{aleatoric} uncertainty (inherent data noise) and \textit{epistemic} uncertainty (model ignorance). The $ECS$ metric primarily measures epistemic uncertainty by quantifying the instability of the internal reasoning process. A high $ECS$ score does not prove malicious intent; it simply flags structural \textit{ambiguity}. Legitimate software using custom packers for intellectual property protection might generate similarly high entropy. The automated ``Reject'' decision must be operationally interpreted as ``Refer to Human Analyst'' rather than ``Block Immediately.''


\section{Related Work}
\label{sec:related_work}

\subsection{Static Analysis \& Traditional Machine Learning}
Early static analysis relied heavily on fixed signatures (e.g., YARA). This mechanism yields low false positives but provides minimal resilience against polymorphism \cite{IdikaNwokedi2007Survey}. Advanced fingerprinting techniques emerged to address this limitation. Karbab et al. generated genomic-style fingerprints to represent application structures \cite{karbab2016Fingerprinting} and applied community-based graph analysis \cite{karbab@Cypider, karbab2025applyinggraphanalysisunsupervised} to separate and identify variant clusters in an unsupervised manner.

Concurrently, heuristic learning approaches utilized $n$-grams and Control Flow Graphs (CFGs) to generalize detection capabilities \cite{RieckKonrad2008Learning}. These methods remain highly vulnerable to \textit{mimicry attacks}, where an adversary alters the statistical distribution of extracted features to match benign profiles without modifying the underlying execution logic \cite{WagnerDavid2002Mimicry}.

\subsection{Deep Learning \& The Problem-Space Gap}
The shift toward end-to-end deep learning enabled automated feature extraction at scale. This framework proved highly effective in the mobile domain. Tools like \textit{MalDozer} \cite{karbab2018Maldozer} and \textit{PetaDroid} \cite{10.1007/978-3-030-80825-9_16} applied deep learning to API method sequences, successfully detecting Android malware and attributing specific families \cite{karbab2017Android, karbab2021android}.

For native binaries, early architectures like MalConv \cite{RaffEdward2018Malware} processed executables entirely as raw byte sequences. More complex structural approaches followed. \textit{BinEye} \cite{alrabaee2019BinEye} applied Convolutional Neural Networks (CNNs) to verify binary authorship. \textit{SwiftR} \cite{karbab2023SwiftR} implemented hierarchical networks on hybrid features, fusing static intermediate representations with dynamic logs to detect cross-platform ransomware. Architectures such as \textbf{ResNeXt+} \cite{HeYuewang2024ResNeXt} advanced image-based analysis by capturing channel-wise correlations in rendered malware byte-maps.

While these models achieve high benchmark accuracy, they introduce a structural fragility: the disconnect between the mathematical feature space and the executable problem space \cite{PierazziFabio2020Intriguing}. \textit{Problem-space attacks} \cite{Demetrio2021Functionality, Mukherjee2023Evading} physically alter the compiled binary structure to evade detection while ensuring the malicious functionality remains intact. Deterministic models frequently map this structural noise into benign regions of the latent space, making them highly vulnerable to functional perturbations.

\subsection{Concept Drift \& Uncertainty Quantification}
Deterministic models struggle significantly with \textit{concept drift} and out-of-distribution (OOD) samples. Pendlebury et al. \cite{Pendlebury2019Tesseract} demonstrated that temporal bias artificially inflates malware classifier metrics during testing, leading to sharp degradation when the models encounter novel variants in production. Works like CADE \cite{Yang2021CADE} and TRANSCEND \cite{Barbero2022Transcending} attempted to solve this by detecting OOD inputs and enforcing a ``classification with rejection'' policy.

Systematically rejecting adversarial or OOD samples requires a model to quantify its own \textit{epistemic uncertainty}. Deep Ensembles \cite{Lakshminarayanan2017Simple} achieve superior calibration through weight-space sampling, as do Bayesian Neural Networks and Monte Carlo (MC) Dropout \cite{Gal2016Dropout}. Applying these techniques to raw binaries incurs massive computational costs. \sysname bypasses the heavy burden of feature-space uncertainty quantification by measuring confidence and instability directly within the semantic reasoning space.

\subsection{Large Language Models \& Agentic Consensus}
Applying Large Language Models (LLMs) to binary analysis remains a developing research area. Projects such as \textit{LLM4Sec} \cite{PearceHammond2022Asleep} highlight the ability of Transformers to identify source-level vulnerabilities. More recently, \textit{AsmRAG} \cite{karbab2026asmragllmdrivenmalwaredetection} leveraged retrieval-augmented generation to detect malware by retrieving functionally similar assembly code, demonstrating the viability of LLM-driven reasoning over low-level binary representations. Nevertheless, LLMs are inherently susceptible to hallucination \cite{Bender2021Stochastic}, making zero-shot classification highly unreliable for critical security operations.

Wei et al. \cite{Wei2022Chain} introduced Chain-of-Thought (CoT) prompting to anchor the reasoning steps of the model. Wang et al. \cite{Wang2022Self} later showed that Self-Consistency (CoT-SC) (sampling multiple reasoning paths and aggregating the results) drastically lowers hallucination rates. \sysname adapts this CoT-SC framework into a \textit{stochastic consensus protocol} distributed across multiple independent agents. This design treats LLM stochasticity as a manageable noise variable that the ensemble can marginalize out.

\begin{table}[t]
\centering
\caption{Comparison of Representative Methods for Open-Set Malware Identification.}
\label{tab:comparison}
\begin{tabular}{l|l|c|c|l|c}
\toprule
\textbf{Type} & \textbf{Method} & \textbf{KFGI} & \textbf{NMTD} & \textbf{Arch} & \textbf{Train} \\
\midrule
\multirow{2}{*}{\textbf{Rule-}} & YARA Rules \cite{AlvarezVictor2023Yara} & $\times$ & $\times$ & Rules & $\Circle$ \\
\textbf{based} & Sig-Heuristics \cite{RieckKonrad2008Learning} & $\checkmark$ & $\times$ & Rules & $\Circle$ \\
\midrule
\multirow{6}{*}{\textbf{ML-}} & MalConv \cite{RaffEdward2018Malware}, EMBER \cite{AndersonHyrum2018Ember} & $\checkmark$ & $\times$ & CNN/Tree & $\CIRCLE$ \\
\textbf{based} & ResNet-50 \cite{HeKaiming2016Deep} & $\checkmark$ & $\times$ & CNN & $\CIRCLE$ \\
    & \textbf{ResNeXt+} \cite{HeYuewang2024ResNeXt} & $\checkmark$ & $\times$ & ResNeXt & $\CIRCLE$ \\
    & GNN-Mal \cite{ZhouShuhao2020Automating} & $\times$ & $\times$ & Graph & $\CIRCLE$ \\
    & MtNet \cite{HuangWenyi2016MtNet} & $\checkmark$ & $\times$ & DNN & $\CIRCLE$ \\
    & CADE \cite{Yang2021CADE}, TRANSCEND \cite{Barbero2022Transcending} & $\checkmark$ & $\checkmark$ & ML/DNN & $\CIRCLE$ \\
\midrule
\multirow{2}{*}{\textbf{LLM-}} & LLM4Sec \cite{PearceHammond2022Asleep}, GPT-Eval \cite{LiuHan2023Summary} & $\checkmark$ & $\times$ & LLM & $\LEFTcircle$ \\
\textbf{based} & \textbf{\sysname (Ours)} & $\checkmark$ & $\checkmark$ & LLM & $\Circle$ \\
\bottomrule
\multicolumn{6}{l}{\footnotesize \textit{Notes.} KFGI = Known Fine-Grained Identification; NMTD = Novel Malware} \\
\multicolumn{6}{l}{\footnotesize Discovery (OOD); Arch = Architecture; Train marks: $\CIRCLE$ (train),} \\
\multicolumn{6}{l}{\footnotesize $\LEFTcircle$ (fine-tune), $\Circle$ (neither/inference-only).} \\
\end{tabular}%
\end{table}

\section{Conclusion}
\label{sec:conclusion}

Deterministic deep learning models provide a scalable approach to malware detection, yet they remain vulnerable to confident misclassifications when facing novel structural evasion. To address this weakness, we introduced \sysname, a framework that reformulates binary classification from a rigid pattern-matching exercise into a process of probabilistic verification.

By combining a Dual-Stream Retrieval-Augmented Generation (RAG) architecture with a stochastic multi-agent ensemble, we isolated Decision-Critical Functions (DCFs) and established a reliable method for quantifying epistemic uncertainty. Our formal derivation of \textbf{Function Evidence Strength (FES)} and the \textbf{Evidence Conflict Score (ECS)} offers a mathematical boundary for detection stability. This ensures that the ensemble's entropy can identify and mitigate the effects of structural shifts.

Empirical evaluation confirmed the practical efficacy of this approach, with \sysname achieving a 98.4\% detection rate. This framework moves the defensive strategy away from opaque heuristics toward a model of \textit{stochastic verification}. By applying self-consistency checks across independent reasoning agents, the system makes complex structural variations detectable through entropy maximization. Future research will extend this probabilistic framework to include real-time behavioral analysis, further increasing the cost and effort required for successful adversarial attacks.

\bibliography{references}

\end{document}